\documentclass[%
 reprint,
 amsmath,amssymb,
 aps,
]{revtex4-2}

\pdfoutput=1 
\usepackage{graphicx}
\usepackage{dcolumn}
\usepackage{bm}
\usepackage[caption=false]{subfig}
\usepackage{amssymb}
\usepackage{amsmath}
\usepackage{commath}
\usepackage{graphicx,bm}
\usepackage{verbatim}

\usepackage[colorlinks=true,linkcolor=blue,citecolor=blue,allcolors=blue]{hyperref}%

\usepackage{color}
    \definecolor{darkgreen}{rgb}{0,0.5,0}
    \definecolor{darkred}{rgb}{0.5,0,0}
    \definecolor{darkblue}{rgb}{0,0,0.6}
    \definecolor{purple}{rgb}{0.4,.2,0.7}
    
\def \dd {\mathrm{d}}

\begin{document}

\preprint{APS/123-QED}

\title{Static Black Binaries in de Sitter}

\author{\'Oscar~J.~C.~Dias}
\email{ojcd1r13@soton.ac.uk}
\affiliation{STAG research centre \& Mathematical Sciences, University of Southampton, Highfield Campus, UK}
\author{Gary~W.~Gibbons}
\email{gwg1@maths.cam.ac.uk}
\affiliation{DAMTP, Centre for Mathematical Sciences, University of Cambridge, Wilberforce Road, Cambridge CB3 0WA, UK}
\author{Jorge~E.~Santos}
\email{jss55@cam.ac.uk}
\affiliation{DAMTP, Centre for Mathematical Sciences, University of Cambridge, Wilberforce Road, Cambridge CB3 0WA, UK}%
\author{Benson~Way}
\email{benson@icc.ub.edu}
\affiliation{Departament de F\'{i}sica Qu\`{a}ntica i Astrof\'{i}sica, Institut de Ci\`{e}ncies del Cosmos Universitat de Barcelona, Mart\'{i} i Franqu\`{e}s, 1, E-08028 Barcelona, Spain}

\begin{abstract}
We construct the first four-dimensional multi-black hole solution of general relativity with a positive cosmological constant. The solution consists of two static black holes whose gravitational attraction is balanced by the cosmic expansion. These static binaries provide the first four-dimensional example of non-uniqueness in general relativity without matter.
\end{abstract}

\maketitle
\subparagraph{Introduction.}
Black holes are famously featureless.  This idea is embodied by the no-hair theorems, which state in essence that stationary black holes are uniquely characterized by their mass, angular momentum, and charge \cite{Chase:1970,Penney:1968zz,Bekenstein:1972ny,Bekenstein:1971hc,Bekenstein:1972ky,Teitelboim:1972qx,Hartle1972,Heusler:1992ss,Bekenstein:1995un,Bekenstein:1996pn,Sudarsky:1995zg}.

It should be noted that there are many situations where black hole uniqueness, as we have expressed it, is known to be violated. A well-known example involves multi-horizon configurations of charged, extremal black holes \cite{Majumdar:1947eu,Papaetrou:1947ib}. Other examples include higher dimensions \cite{Emparan:2001wn}, anti-de Sitter asymptotics \cite{Gubser:2008px,Hartnoll:2008kx,Dias:2011at}, or exotic matter like classical Yang-Mills fields, complex scalars and Proca fields \cite{Volkov:1998cc,Herdeiro:2014goa,Herdeiro:2016tmi}.

Additionally, there are some mathematical gaps in fully establishing black hole uniqueness, even in the more limited case of four-dimensional pure gravity in flat space.  Indeed, asymptotically flat multi-Kerr black holes, where their gravitational attraction might be balanced by spin-spin interactions, have not been ruled out (see \emph{e.g.} 
\cite{PerjesPhysRevLett.27.1668,Israel:1972vx,Hartle:1972ya,Tomimatsu:1972zz,KRAMER1980259,NEUGEBAUER198191,stephani_kramer_maccallum_hoenselaers_herlt_2003,Manko2001,Herdeiro:2008kq,Manko:2008pv,Manko:2017avt,Manko:2018iyn,Manko:2020jfa} for attempted constructions that yield singular configurations). Though for static solutions, a classic theorem due to \cite{Israel:1967wq,1977GReGr...8..695R,1987GReGr..19..147B} precludes the existence of regular asymptotically flat multi black holes.

Despite these (and potentially more) counterexamples, there is currently no experimental or observational evidence that black hole non-uniqueness can be realized in astrophysical or cosmological contexts. Indeed, the no-hair theorems are fully consistent with observational results from the LIGO consortium \cite{KAGRA:2013rdx}. 

However, the no-hair theorems assume that spacetime is asymptotically flat, a feature which is violated in our universe at the longest scales by the presence of a cosmological constant \cite{SupernovaCosmologyProject:1997zqe,SupernovaSearchTeam:1998fmf,SupernovaCosmologyProject:1998vns,Wright2011ER,Planck:2013pxb}.  The resulting cosmic expansion might balance out the gravitational attraction of two or more black holes, allowing multiple black holes to exist in static equilibrium. Such a configuration would share the same mass and angular momentum as some single-horizon black hole and therefore serve as a more realistic counterexample to black hole uniqueness.

The aim of this Letter is to demonstrate that such a multi-horizon configuration indeed occurs within general relativity.  We will focus on the simplest case with two equal-mass black holes that do not rotate nor contain charge, but our results and methods can be straightforwardly generalized.  We will first show that these black binaries can be anticipated using intuition from Newton-Hooke theory, and then construct these solutions by solving the Einstein equation numerically. Finally, we study the properties of these binaries in detail.  

Our results, along with physical intuition, suggest that the static de Sitter binaries are dynamically unstable.  Nevertheless, there remains a possibility that they can be stabilized with the introduction of charge or angular momentum.  We will comment on this and other matters in the conclusions.

Before we continue, we mention some closely related work.  Dynamical (\emph{i.e.} out of equilibrium) multi-black holes in Einstein-Maxwell theory with a positive cosmological constant were found in \cite{Kastor:1992nn}. The ``rod-structure" corresponding to our static binaries were anticipated and examined in detail in \cite{Armas:2011ed}.  In \cite{Astorino:2022fge}, a novel mechanism for balancing multi-black holes was proposed. These constructions provide Ricci-flat, closed-form solutions for static binaries supported by expanding bubbles of nothing. Mechanically, these solutions behave similarly to the static binaries we find.

Finally, we mention the mathematical papers \cite{LEFLOCH20101129,Borghini:2019msu,masood-ul-alam_yu_2014}, which might seem to rule out the existence of static black binaries in de Sitter.  We will show that the assumptions made in \cite{LEFLOCH20101129,Borghini:2019msu} do not apply, and that (for technical reasons) this conclusion from \cite{masood-ul-alam_yu_2014} is not correct. 

\subparagraph{Newton-Hooke.}
Let us first set out to see if the aforementioned multi-black hole configurations are allowed within Newtonian gravity.  We adopt geometrized units in which $c=G=k_B=\hbar=1$.

Consider a configuration of $N$ black holes with masses $m_a$, with $a=1,\ldots,N$.  For the Newtonian approximation to be valid, we assume that the distances between the black holes are much larger than their masses.  We now include the effects of the cosmological constant $\Lambda\equiv 3/\ell^2>0$, where $\ell$ is the de Sitter length scale.  Accordingly, we assume that the entire configuration of black holes lies within a distance much smaller than $\ell$ and consider the Newton-Hooke equations of motion \cite{Battye:2002gn,Gibbons:2003rv}
\begin{equation}
m_a \frac{\mathrm{d}^2\mathbf{x}_a}{\mathrm{d}t^2}-m_a \frac{\mathbf{x}_a}{\ell^2}=-\sum_{b\neq a}^{b=N} \frac{\,m_a\,m_b(\mathbf{x}_a-\mathbf{x}_b)}{|\mathbf{x}_a-\mathbf{x}_b|^3}\,,
\end{equation}
where $\mathbf x_a$ are the positions of the black holes.

Static solutions exist when
\begin{equation}
\frac{\mathbf{x}_a}{\ell^2}=\sum_{b\neq a}^{b=N} \frac{\,m_b(\mathbf{x}_a-\mathbf{x}_b)}{|\mathbf{x}_a-\mathbf{x}_b|^3}\,.
\label{eq:central}
\end{equation}
Such solutions are known  as {\it central configurations} and provide homothetic solutions of the Newtonian $N$-body problem which has applications to Newtonian cosmology. The equation \eqref{eq:central} can also be obtained from the Dmitriev-Zeldovich equations \cite{dmitriev1965semi} by using the scale factor $S(t) = e^{\frac{t}{\ell} }$, corresponding to a de Sitter background in ``stead-state" coordinates \cite{Battye:2002gn,Gibbons:2003rv,Ellis:2013xjx,Ellis:2014sla}. 

Consider a central configuration with two equal mass black holes aligned along the $z$ axis and separated by a distance $d$.  That is, $N=2$, $\mathbf{x}_1=-\mathbf{x}_2=\frac{d}{2}\,\mathbf{e}_z$, and $m_a=m_b=M$.  Then \eqref{eq:central} imposes
\begin{equation}
\frac{d^3}{\ell^3}=\frac{r_+}{\ell}
\label{eq:pred}
\end{equation}
where $r_+\equiv 2M$ is the Schwarzschild radius.  

The requirements that the Newton-Hooke approximation should be valid and that the black holes are inside a single cosmological event horizon amount to
\begin{equation}
r_+\ll d\,, \quad d\ll \ell\quad\text{and}\quad r_+\ll \ell\,.
\end{equation}
If the distance between the black holes is given as in \eqref{eq:pred}, then we see that the first two conditions above are satisfied if we assume the third, \emph{i.e.} if the black holes are small enough.  We therefore conclude that static de Sitter binaries with small black holes are consistent with Newton-Hooke theory. 

For later use, we introduce the event horizon Hawking temperature $T_+= (4 \pi r_+)^{-1}$ and rewrite \eqref{eq:pred} as
\begin{equation}
\frac{d}{\ell} = \frac{1}{(4 \pi \ell \,T_+)^{1/3}}\,.
\label{eq:pred2}
\end{equation}
We will confirm that our numerical solutions to the Einstein equation satisfy this scaling in the appropriate limit. 

\subparagraph{Numerical construction.} 
We now construct static binaries in general relativity by numerically solving the Einstein equation with a positive cosmological constant:
\begin{equation}
R_{ab}=\frac{3}{\ell^2}\,g_{ab}\,,
\label{eq:einstein}
\end{equation}
where $R_{ab}$ is the Ricci tensor and $g_{ab}$ is the metric tensor. 

We use the DeTurck method, first introduced for general relativity in \cite{Headrick:2009pv} and reviewed in \cite{Wiseman:2011by,Dias:2015nua}.  This method provides a convenient way of addressing the issue of gauge invariance, which ultimately causes the Einstein equation \eqref{eq:einstein} to yield a set of ill-posed, non-elliptic PDEs. 

The DeTurck method involves choosing any reference metric $\bar{g}$ with the same symmetries and causal structure as the solution we seek.  In this case, our reference metric is static, contains two identical black holes, a cosmological horizon, and is axisymmetric.  There is therefore a discrete $\mathbb Z_2$ symmetry, as well as two Killing vector fields $k=\partial/\partial t$ and $m=\partial/\partial \phi$. We further assume that the black holes and cosmological horizon are Killing horizons generated by $k$.  Our choice of reference metric involves a combination of the Bach-Weyl solution with two identical black holes \cite{BachWeyl(Republication):1922} (equivalent to the Israel-Khan solution \cite{Israel1964,Emparan:2001bb} with two black holes) and the static patch of de Sitter space.  Its design is detailed in the Supplementary Material. 

We then write down the most general metric ansatz $g$ that respects the desired symmetries and causal structure.  In this case, the metric ansatz depends non-trivially on two coordinates (\emph{i.e.} it is cohomogeneity-two, and will yield two-dimensional PDEs).  

We then solve the Einstein-DeTurck equation
\begin{equation}
R_{ab}-\nabla_{(a}\xi_{b)}=\frac{3}{\ell^2}g_{ab}\,,
\label{eq:deturck}
\end{equation}
where $\xi^a \equiv g^{bc}\left[\Gamma^a_{bc}(g)-\Gamma^a_{bc}(\bar{g})\right]$, and $\Gamma(\mathfrak{g})$ is the metric-preserving Christoffel connection associated to a metric $\mathfrak{g}$.  Unlike the Einstein equation, the Einstein-DeTurck equation \eqref{eq:deturck} yields a set of elliptic PDEs \cite{Headrick:2009pv,Wiseman:2011by,Figueras:2011va,Dias:2015nua}, which gives a well-posed boundary-value problem with appropriate physical boundary conditions.  

The Einstein-DeTurck equation \eqref{eq:deturck} is solved numerically.  One complication is that the integration domain contains five boundaries: the $\mathbb Z_2$ reflection surface, the inner and outer portions of the symmetry axes, the black hole horizons, and the cosmological horizon.  We handle this domain using patching techniques.  This and other numerical methods we use are described in \cite{Dias:2015nua} and detailed in the Supplementary Material.

After solving \eqref{eq:deturck}, we must verify that the solution actually solves the Einstein equation, \emph{i.e.} that $\xi=0$, and is therefore not a Ricci soliton (for which $\xi\neq0$).  Under many circumstances \cite{Figueras:2011va,Figueras:2016nmo}, it can be proved that these unwanted Ricci solitons do not exist.  Unfortunately, the present case is not one of these circumstances. Indeed, with a positive cosmological constant, Ricci solitons are known to exist (see \emph{e.g.} \cite{10.4310/jdg/1090511686}). Nevertheless, ellipticity guarantees local uniqueness. That is, solutions with $\xi=0$ cannot be arbitrarily close to solutions with $\xi\neq0$, and thus the norm $\xi^a\xi_a$ can be monitored to identify whether our numerical discretization converges in the continuum to a Ricci soliton or to a true solution of the Einstein equation. In the Supplementary Material, we provide ample evidence that the numerical solutions we construct are \emph{not} Ricci solitons.

\subparagraph{Results.} Having numerical solutions corresponding to static black binaries in de Sitter, we can now describe their properties and compare the numerical results to Newton-Hooke theory when the latter is valid.  

\begin{figure}
    \centering
    \includegraphics[width=0.4\textwidth]{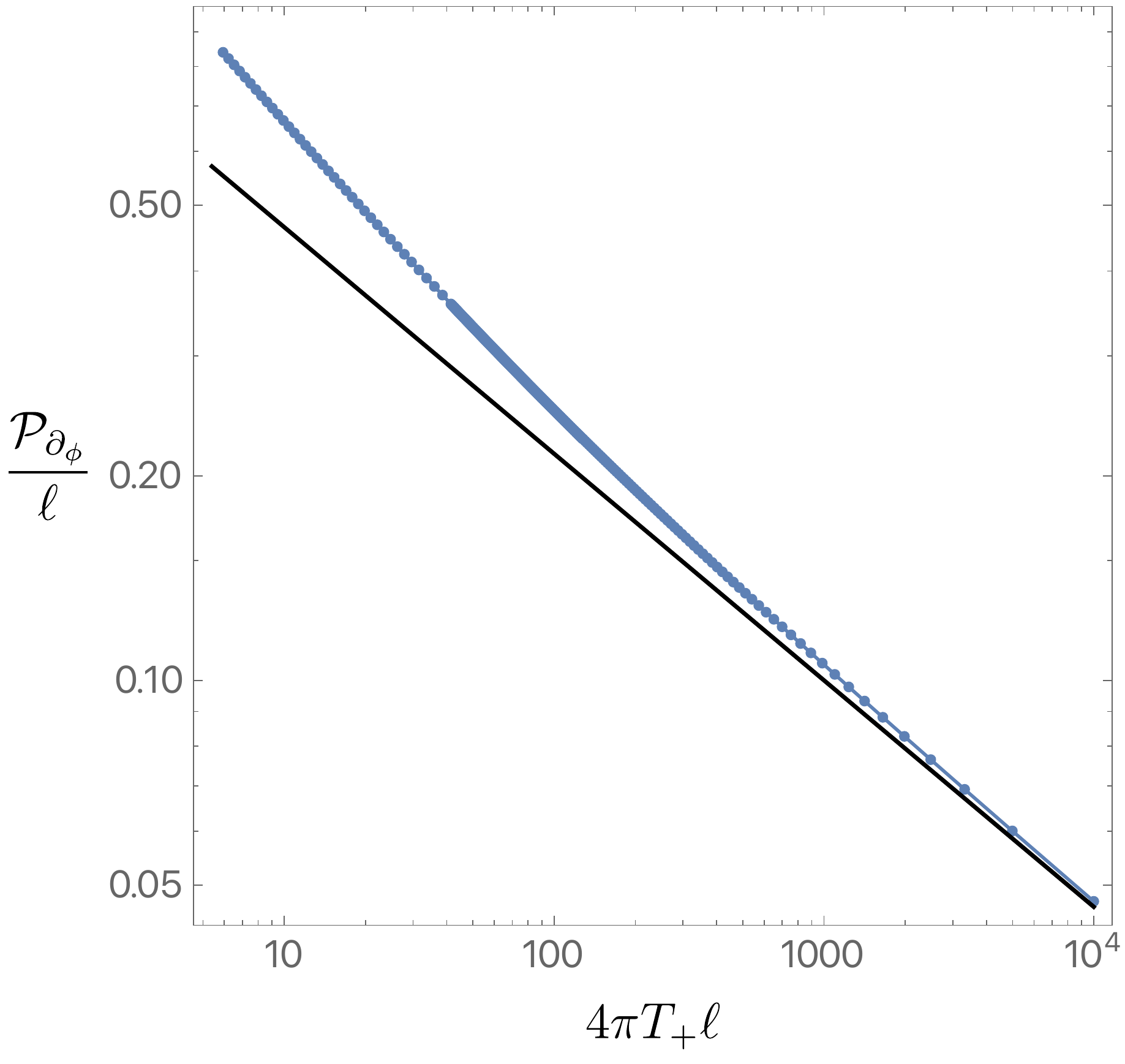}
    \caption{Proper distance between the black hole horizons versus the black hole temperature.  The solid black line shows the scaling \eqref{eq:pred2} according to Newton-Hooke analysis and the blue dots show numerical data according to general relativity.}
    \label{fig:proper_distance}
\end{figure}

We expect to find agreement with Newton-Hooke theory when the black holes become sufficiently small, or alternatively, when the black hole temperature becomes sufficiently large $T_+ \ell\gg 1$. In FIG.~\ref{fig:proper_distance}, we provide a log-log plot of the proper distance between the horizons of the two black holes along the symmetry axis $\mathcal{P}_{\phi}/\ell$, as a function of temperature $4 \pi T_{+} \ell$. The solid black line is the scaling \eqref{eq:pred2}, and the blue dots are the numerical data. The agreement at large values of $T_+\ell$ shows the validity of the Newton-Hooke analysis and corroborates our numerical construction.  

We have not managed to find solutions with large black holes (small $4\pi T_+ \ell$).  Because our solutions do not have regions of large curvature, there might be a ``turning point" to a new branch of solutions. A similar phenomenon occurs for localized Kaluza-Klein black holes when the black holes are large relative to the Kaluza-Klein circle \cite{Kol:2002xz,Wiseman:2002zc,Kol:2003ja,Harmark:2003yz,Gorbonos:2004uc,Harmark:2004ws,Asnin:2006ip,Harmark:2002tr,Wiseman:2002ti,Kudoh:2003ki,Kudoh:2004hs,Sorkin:2006wp,Kleihaus:2006ee,Harmark:2007md,Dias:2007hg,Headrick:2009pv,Wiseman:2011by,Figueras:2012xj,HorowitzBook2012,Kalisch:2016fkm,Dias:2016eto,Dias:2017uyv}.  We leave the exploration of this region of parameter space for future work. 

Let us now discuss black hole thermodynamics. For a central configuration containing $N$ black holes inside the static patch of de Sitter, the covariant phase space formalism \cite{wald1993black, iyer1994some, iyer1995comparison, wald2000general,papadimitriou2005thermodynamics,Anderson:1996sc,Barnich:2001jy,Barnich:2003xg,Barnich:2007bf,Compere:2007vx,Chow:2013gba,Compere:2007az} shows that the following form of the first law of black hole mechanics holds
\begin{equation}
\sum_{i=1 }^{N}T^{(i)}_+ \,\mathrm{d}S^{(i)}_+ = -T_c\,\mathrm{d}S_c\,,
\end{equation}
where $T_c$ is the temperature of the cosmological horizon, and $S_c$ is its entropy (\emph{i.e} horizon area). $T^{(i)}_+$ and $S_+^{(i)}$ are the same quantities, respectively, for the $i-$th black hole.  With $N=2$ and equal-mass black holes, we find
\begin{equation}
2 T_+ \,\mathrm{d}S_+ = -T_c\,\mathrm{d}S_c\,.
\end{equation}
We have checked that our data satisfies this form of the first law to within $0.01\%$.

\begin{figure}
    \centering
    \includegraphics[width=0.4\textwidth]{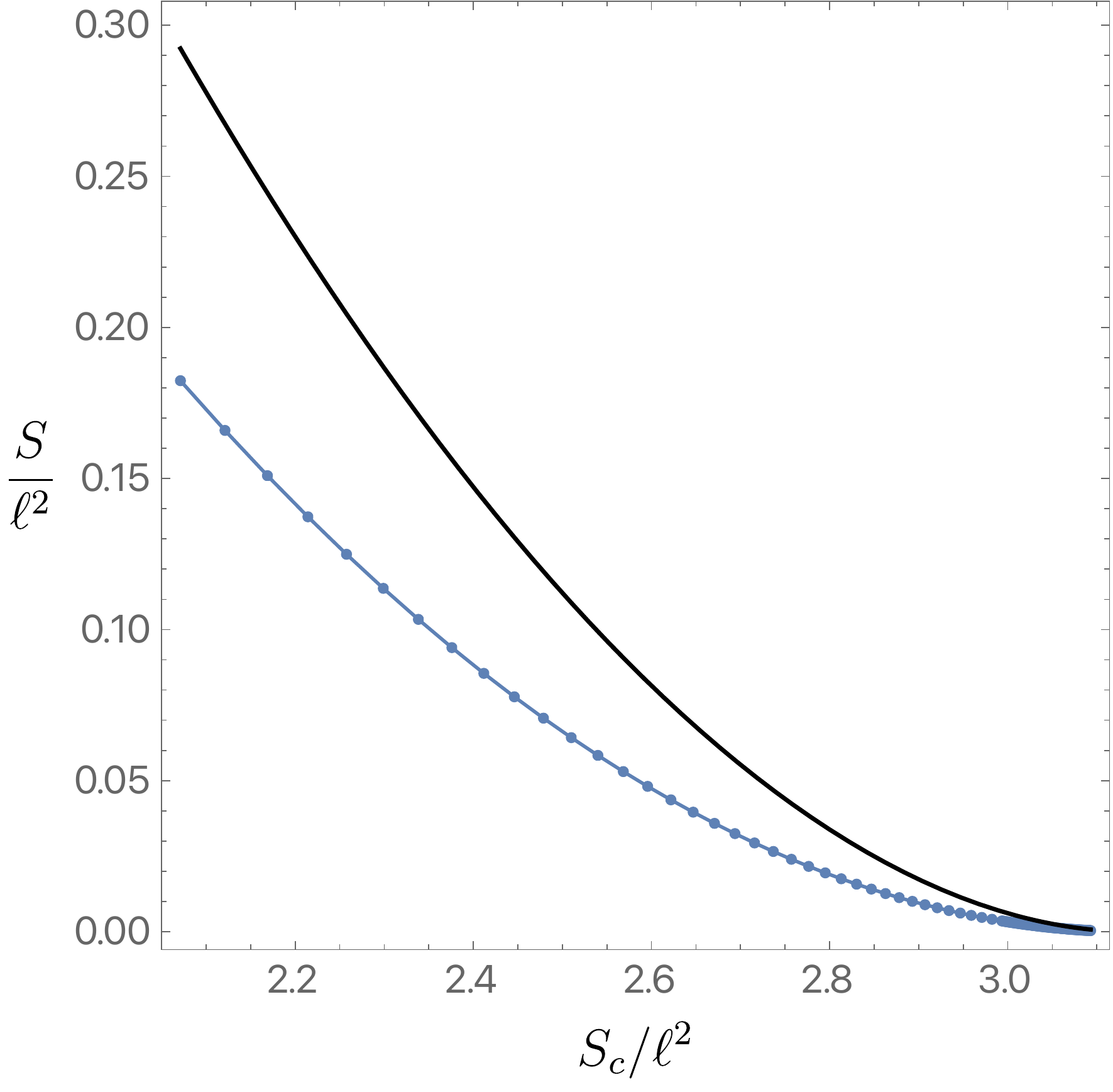}
    \caption{Total black hole entropy versus the cosmological horizon entropy. The blue dots are numerical data for static binaries ($S=2S_+$) and the solid black line is for the single Schwarzschild-de Sitter black hole.}
    \label{fig:micro}
\end{figure}

Following \cite{Gibbons:1977mu}, we now consider the entropy, which must increase during time evolution.  The blue dots in FIG.~\ref{fig:micro} show the entropy of the static binary as a function of the entropy $S_c$ of the cosmological horizon.  The black curve shows the entropy for the single Schwarzschild-de Sitter black hole (also known as the Kottler black hole).  We see that for any given $S_c$, the single Schwarzschild black hole has higher entropy than the binary.  This, together with the second law of thermodynamics, indicates that classically the binary can evolve towards the single black hole but not the other way around. The static black binary is therefore thermodynamically unstable. Beyond the classical level, when the black holes are small, Hawking radiation should however also play a role in this discussion.

The fact that (at least) two solutions exist for a given cosmological horizon entropy implies that the Schwarzschild-de Sitter black hole is not unique.  This is the first counterexample to the no-hair conjecture \cite{Ruffini:1971bza} for pure gravity with a positive cosmological constant. 

\begin{figure*}
    \centering
    \includegraphics[width=0.8\textwidth]{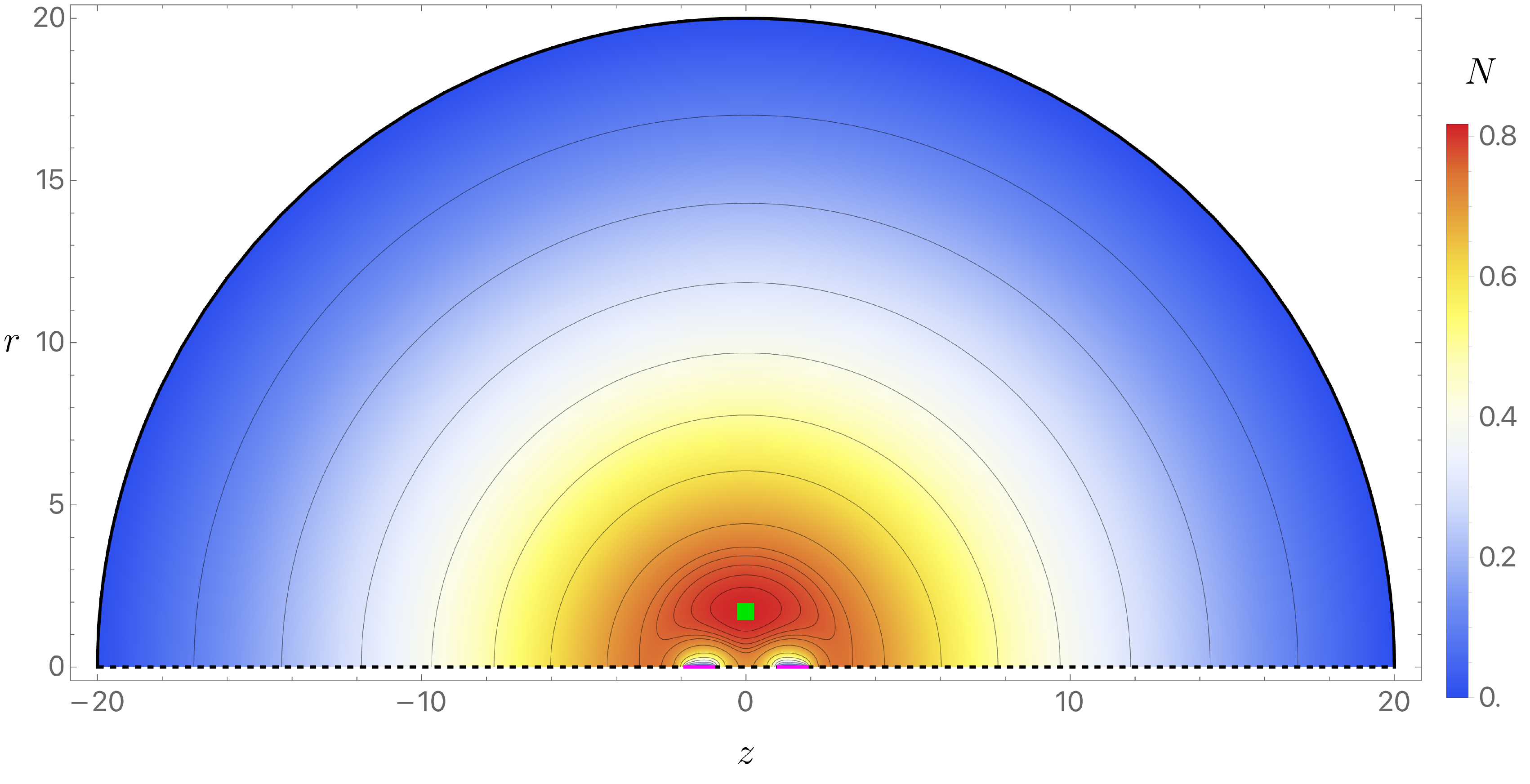}
    \caption{Contour plot showing the level sets of the lapse function $N$.  The cosmological horizon is the outer solid black semicircle.  The horizon axes has the two black hole horizons as solid magenta lines, and the outer and inner axes in dashed black lines. The green square is where $N$ takes its maximum value.}
    \label{fig:level}
\end{figure*}

We now comment on the uniqueness theorems for de Sitter black holes \cite{LEFLOCH20101129,Borghini:2019msu,masood-ul-alam_yu_2014} which would, under certain assumptions, rule out the existence of static de Sitter binaries. In \cite{LEFLOCH20101129}, the level sets of the lapse function $N\equiv \sqrt{-g_{tt}}$ are assumed to be surface forming.  In particular, this means that the level sets must consist only of 2D surfaces. In \cite{Borghini:2019msu}, the set $\mathrm{MAX}(N)=\{x\in\mathcal{M}:N(x)=N_{\max}\}$, where $N_{\max}$ is the maximum value of $N$ in the manifold $\mathcal{M}$, is assumed to disconnect $\mathcal{M}$ into an inner region $\mathcal{M}_-$ and an outer region $\mathcal{M}_+$ with the same virtual mass.  Our static binaries do not satisfy either of these assumptions.  Indeed, in FIG.~\ref{fig:level} we show the level sets $N$ in our domain of integration for a typical solution (all of our solutions show the same qualitative behavior). The coordinates $(r,z)$ are defined in the Supplemental Material.  The cosmological horizon is represented by the outer solid black semicircle, the two black hole horizons are marked by solid magenta lines along the line $r=0$, and the inner and outer portions of the symmetry axes are given by the dashed horizontal line. Finally, the green square marks the location of the maximum of $N$ in $\mathcal{M}$.  This maximum represents an $S^1$ on the manifold, which is not a 2D surface, and it also does not partition the manifold into two regions. Therefore, our static binaries fail to satisfy the assumptions in \cite{LEFLOCH20101129,Borghini:2019msu}.

Finally, we comment on \cite{masood-ul-alam_yu_2014}. We believe that this work is not correct for a rather technical reason. Beginning with the Schwarzschild-de Sitter black hole, the authors in \cite{masood-ul-alam_yu_2014} argue that they can construct an asymptotically flat metric that is conformal to the original one, is topologically $S^1\times S^2$ deprived of one point, and has zero ADM mass.  If that were true, the rigidity statement in the positive mass theorem  \cite{Schon:1979rg,Schon:1981vd,Witten:1981mf,Gibbons:1982jg,Lee:2021kgg} would not only imply that the original metric is conformally flat, but also that $S^1\times S^2$ with one point removed is diffeomorphic to $\mathbb{R}^3$, which is impossible.

\subparagraph{Conclusions.} We constructed the first example of a multi-black hole solution within general relativity with a positive cosmological constant and established that the leading behavior of these solutions agrees (for small black holes) with estimates from Newton-Hooke theory. Based on thermodynamic considerations, we argued that these solutions are thermodynamically unstable.  Because the configuration requires a delicate balance between gravitational attraction and cosmic expansion, we expect these solutions to also be dynamically unstable. 

We have focused on the static configuration of two identical black holes, but our results and methods can be generalized.  First, consider the case where the black holes have different masses.  When one of the black holes is much smaller than the other, one can use the geodesic approximation to predict the existence of such a configuration. Indeed, one can easily confirm the existence of static orbits for timelike particles on a Schwarzschild-de Sitter black hole background, thus providing further evidence for the existence of this more general central configuration. Note that if \cite{masood-ul-alam_yu_2014} were correct, this asymmetric binary would also not exist.

We can also include rotation, which will introduce spin-spin interaction of the black holes.  This opens the possibility of continuous non-uniqueness.  Consider, for example, the case with two identical black holes rotating in opposite directions along the axis of symmetry.  This configuration will have vanishing total angular momentum, and will thus be in the same class as the Schwarzschild-de Sitter black hole. Work in this direction is underway.

Perhaps more interestingly, because spin-spin interactions act on shorter length scales, they could provide a mechanism for stabilizing the binary. This possibility resembles the mechanism that provides stability for molecules.  Work in this direction is underway.

We could also consider central configurations containing $N>2$ static black holes in the static patch of de Sitter.  These configurations can show interesting properties within the Newton-Hooke approximation.  For instance, when $N\geq13$, minimal energy central configurations do not lie on a regular polyhedron \cite{Battye:2002gn}.  We thus expect the equivalent property within general relativity.  The study of these configurations is within the reach of the numerical methods employed in this Letter.
\newpage
\begin{acknowledgments}
The authors would like to thank Stefano~Borghini, Lorenzo~Mazzieri and Ryan~Unger for valuable discussions and correspondence regarding \cite{masood-ul-alam_yu_2014}. J.~E.~S. would like to thank Jay~Armas and Troels~Harmark for mentioning this interesting problem to him in 2012 at the 28th Pacific Coast Gravity Meeting, held at University of California, Santa Barbara and Don~Marolf for discussions on the microcanonical properties of black holes contained inside the static patch of de Sitter.  O.~C.~D. acknowledges financial support from the STFC ``Particle Physics Grants Panel (PPGP) 2018" Grant No.~ST/T000775/1. O.D.'s research was also supported in part by the International Centre for Theoretical Sciences (ICTS), India, in association with the program "Nonperturbative and Numerical Approaches to Quantum Gravity, String Theory and Holography " (code: ICTS/numstrings-2022/8). J.~E.~S. has been partially supported by STFC consolidated grant ST/T000694/1.  B.~W. would like to thank Roberto Emparan for fruitful discussions and acknowledges support from ERC Advanced Grant GravBHs-692951 and MEC grant FPA2016-76005-C2-2-P.  The authors acknowledge the use of the IRIDIS High Performance Computing Facility, and associated support services at the University of Southampton, for the completion of this work.
\end{acknowledgments}


\widetext

\appendix

\section*{Supplementary Material}


\section{The Bach-Weyl (Israel-Khan) Solution}
The Bach-Weyl (1922) solution \cite{BachWeyl(Republication):1922} is an exact solution that describes two asymptotically flat black holes separated by a conical strut (the perhaps more familiar Israel-Khan (1964) solution \cite{Israel1964} reduces to it and generalizes it to describe a system with $N\geq2$ black holes). We will be interested in the case where both black holes are equal.  Because it is an exact multi-horizon solution, it will be useful for our numerical construction of static de Sitter binaries. Here, we review this solution and describe some of its coordinate representations that we use. 

Aside from the conical strut, the Bach-Weyl (Israel-Khan) spacetime is completely regular outside the horizons. Here, we are only concerned with the case where both black holes are equal.  The solution is often presented in Weyl coordinates (see e.g. \cite{BachWeyl(Republication):1922,Israel1964,Emparan:2001bb}):
\begin{equation}
\dd s^2=\ell^2\left[-f\dd t^2+\frac{\lambda^2}{f}[h(\dd r^2+\dd z^2)+r^2\dd \phi^2]\right]\;,
\end{equation}
where
\begin{equation}
f=\left(\frac{k(R_++r_+)-(1-k)}{k(R_++r_+)+(1-k)}\right)\left(\frac{k(R_-+r_-)-(1-k)}{k(R_-+r_-)+(1-k)}\right)\;,
\end{equation}
\begin{align}
h&=\left(\frac{r^2+(z+1/k)(z+1)+R_+r_+}{2R_+r_+}\right)\left(\frac{r^2+(z-1/k)(z-1)+R_-r_-}{2R_-r_-}\right)\nonumber\\
&\qquad \times\left(\frac{r^2+z^2-1+r_+r_-}{r^2+(z-1/k)(z+1)+r_+R_-}\right)\left(\frac{r^2+z^2-(1/k^2)+R_+R_-}{r^2+(z+1/k)(z-1)+R_+r_-}\right)\;,
\end{align}
with
\begin{equation}
R_\pm=\sqrt{r^2+\left(z\pm\frac{1}{k}\right)^2}\;,\qquad r_\pm=\sqrt{r^2+(z\pm1)^2}\;,
\end{equation}
and where $\ell$ is an arbitrary length scale that we have introduced for later use in de Sitter. The solution is parametrized by $\lambda\in(0,\infty)$ and $k\in(0,1)$.  The temperature of the black holes is given by
\begin{equation}\label{IKtemp}
    T_+ = \frac{1}{2\pi}\frac{k(1+k)}{4\lambda(1-k)}\,.
\end{equation}

A peculiarity of the Weyl form is that the axis and horizons of the solution are all located at $r=0$, and so is described as a ``rod structure."  At $r=0$, the horizons lie in the regions $z\in (1,1/k)$ and $z\in (-1/k,-1)$, with the inner segment of the axis between the black holes in the region $z\in(-1,1)$, and the outer segments of the axis in $z\in(1/k,\infty)$ and $z\in(-\infty,-1/k)$.  The inner segment of the axis contains a conical singularity which holds the two black holes apart.

The Weyl form is useful for obtaining this solution as it provides a means of simplifying the Einstein equation into an integrable form.  But in order to accommodate the rod structure, the coordinates cannot be smooth along the line $r=0$ (there are coordinate singularities at the ``joints" between rods), making the Weyl form ill-suited for our numerical purposes.

\begin{figure}[ht]
\centering
\includegraphics[width=0.3\textwidth]{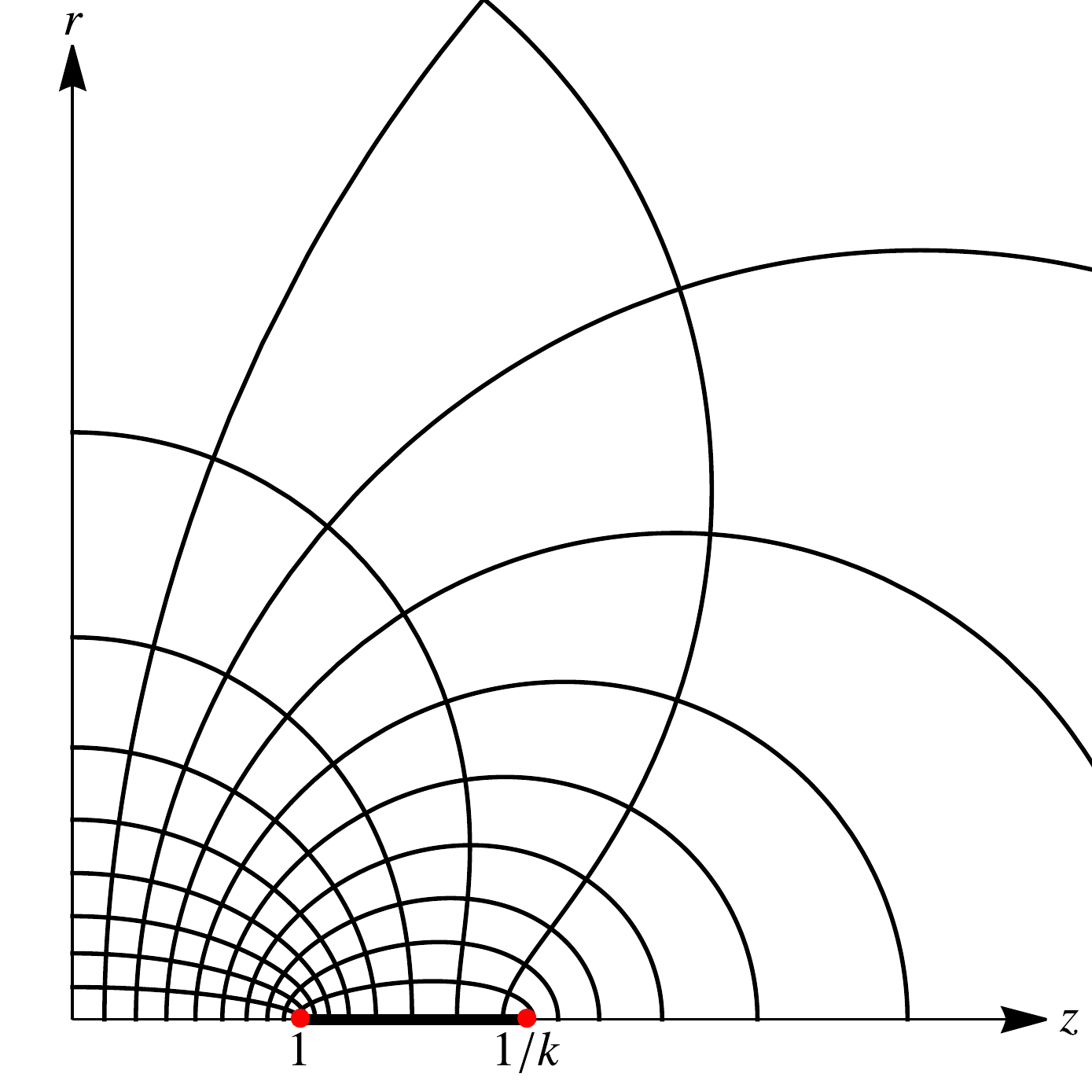}
\caption{Lines of constant $x$ and $y$, shown in Weyl coordinates.  There is a reflection symmetry about $z=0$, so only positive $z$ is shown. The rod structure of the Bach-Weyl (Israel-Khan) solution can be seen at $r=0$, with one of the black hole horizons lying between the two red dots.}\label{fig:xy}
\end{figure}

We therefore seek a coordinate transformation that maps the outer and inner segments of the axis, and horizons into a coordinate rectangle.  This can be accomplished by a conformal Schwarz-Christoffel transformation.  The standard formulas for this type of transformation will give mappings that use Jacobi Elliptic functions, but here we convert these functions to a more algebraic form.  The mappings we use are defined by
\begin{equation}\label{xydef}
z=\frac{x\sqrt{2-x^2}\sqrt{(1-y^2)^2+k^2y^2(2-y^2)}}{(1-y^2)^2+k^2x^2(2-x^2)y^2(2-y^2)}\;,\qquad r=\frac{(1-x^2)\sqrt{1-k^2x^2(2-x^2)}y\sqrt{2-y^2}(1-y^2)}{(1-y^2)^2+k^2x^2(2-x^2)y^2(2-y^2)}\;.
\end{equation}
Lines of constant $x$ and $y$, along with the rod structure of Bach-Weyl are shown in FIG.~\ref{fig:xy}.

With this coordinate transformation, the Bach-Weyl solution takes the form
\begin{equation}
\label{IKxy}
\dd s^2=\ell^2\left\{-f\dd t^2+\frac{\lambda^2}{m^2 \Delta_{xy}^2}\left[p^2\left(\frac{4\dd x^2}{(2-x^2)\Delta_x}+\frac{4\dd y^2}{(2-y^2)\Delta_y}\right)+y^2(2-y^2)(1-y^2)^2\dd \phi^2\right]\right\}\;,
\end{equation}
where in these coordinates
\begin{equation}
f=(1-x^2)^2\Delta_x m^2
\end{equation}
and
\begin{equation}
\Delta_x=1-k^2x^2(2-x^2)\;,\qquad \Delta_y=1-(1-k^2)y^2(2-y^2)\;,\qquad \Delta_{xy}=(1-y^2)^2+k^2x^2(2-x^2)y^2(2-y^2)\;,
\end{equation}
\begin{equation}
p=\frac{k}{(1+k)^2}\left(1+\sqrt{\Delta_y}\right)^2\;,\qquad m=\frac{k\Big[1-(1-k)y^2(2-y^2)+\sqrt{\Delta_y}\Big]}{(1-k)\Delta_x(1-y^2)^2+(k+\sqrt{\Delta_y})\Big[\Delta_x+(1-k)(\sqrt{\Delta_{xy}}-1)\Big]}\;.
\end{equation}
The horizons are at $x=\pm 1$, the inner segment of the axis is at $y=0$, and the outer segments of the axis are at $y=1$.  There is also a $\mathbb Z_2$ symmetry about $x=0$.  All functions $\Delta_x$, $\Delta_y$, $w$, and $m$ are smooth and positive definite in the domain.  $\Delta_{xy}$ vanishes at $(x,y)=(0,1)$ (asymptotic infinity), and is positive and smooth otherwise.

Eventually, we wish to join the Bach-Weyl solution with a de Sitter horizon.  In anticipation of doing so, we present the Bach-Weyl solution in polar-Weyl coordinates defined by
\begin{equation}\label{rhoxidef}
z=\rho\,\xi\sqrt{2-\xi^2}\;,\qquad r=\rho(1-\xi^2)\;,
\end{equation}
where the Bach-Weyl (Israel-Khan) solution takes the form
\begin{equation}\label{IKpolar}
\dd s^2=\ell^2\left\{-f\dd t^2+\frac{\lambda^2h}{f}\left[\dd \rho^2+\rho^2\left(\frac{4\dd\xi^2}{2-\xi^2}+\frac{(1-\xi^2)^2}{h}\dd\phi^2\right)\right]\right\}\;,
\end{equation}
with $f$ and $h$ also transformed accordingly.  For later use, we find it convenient to express $f$ and $h$ as functions of $\rho$ and $z=\rho\,\xi\sqrt{2-\xi^2}$:
\begin{equation}\label{frhoz}
f=\left(\frac{k(R_++r_+)-(1-k)}{k(R_++r_+)+(1-k)}\right)\left(\frac{k(R_-+r_-)-(1-k)}{k(R_-+r_-)+(1-k)}\right)\;,
\end{equation}
\begin{align}\label{hrhoz}
h&=\left(\frac{\rho^2+\tfrac{1}{k}[1+(1+k)z]+R_+r_+}{2R_+r_+}\right)\left(\frac{\rho^2+\tfrac{1}{k}[1-(1+k)z]+R_-r_-}{2R_-r_-}\right)\nonumber\\
&\qquad \times\left(\frac{\rho^2-1+r_+r_-}{\rho^2-\tfrac{1}{k}[1+(1-k)z]+r_+R_-}\right)\left(\frac{\rho^2-(1/k^2)+R_+R_-}{\rho^2-\tfrac{1}{k}[1-(1-kz)]+R_+r_-}\right)\;,
\end{align}
with
\begin{equation}
R_\pm=\sqrt{\rho^2+\frac{1}{k^2}\pm\frac{2z}{k}}\;,\qquad r_\pm=\sqrt{\rho^2+1\pm2z}\;,
\end{equation}
Note that $h$ and $f$ approach unity when $\rho\to\infty$, where the spacetime becomes asymptotically flat.

From \eqref{xydef} and \eqref{rhoxidef}, we can derive an explicit coordinate transformation between the Schwarz-Christoffel $(x,y)$ coordinates and the polar-Weyl $(\rho,\xi)$ coordinates:
\begin{subequations}\label{trans}
\begin{align}
\rho&=\frac{\sqrt{y^2(2-y^2)+x^2(2-x^2)(1-y^2)^2}}{\sqrt{(1-y^2)^2+k^2x^2(2-x^2)y^2(2-y^2)}}\;,\\
\xi&=\sqrt{1-\frac{(1-x^2)y\sqrt{2-y^2}(1-y^2)\sqrt{1-k^2x^2(2-x^2)}}{\sqrt{y^2(2-y^2)+x^2(2-x^2)(1-y^2)^2}\sqrt{(1-y^2)^2+k^2x^2(2-x^2)y^2(2-y^2)}}}\;.
\end{align}
\end{subequations}
Like the original Weyl coordinates, these polar-Weyl coordinates are not smooth along $\xi=\pm 1$.  This will not be an issue for us as we will only use these coordinates in our numerical construction for sufficiently large $\rho$, where the coordinates are smooth.

\section{Designing the Reference Metric for Einstein-DeTurck}
Our strategy for designing a reference metric for the Einstein-DeTurck problem (described in the main text) is to attach a de Sitter horizon to the Bach-Weyl (Israel-Khan) solution.  De Sitter space in four dimensions is most commonly written in the form
\begin{equation}
\dd s^2=-\left(1-\frac{R^2}{\ell^2}\right)\dd \tau^2+\frac{\dd R^2}{1-\frac{R^2}{\ell^2}}+r^2(\dd\theta^2+\sin^2\theta\dd\phi^2)\;,
\end{equation}
where $\ell$ is the de Sitter length scale.  De Sitter can also be written in isotropic coordinates with the transformation
\begin{equation}
\frac{R}{\ell}=\frac{\lambda\,\rho}{1+\frac{\lambda^2\rho^2}{4}}\;,\qquad \sin\theta=1-\xi^2\;,\qquad \tau=\ell\, t\;,
\end{equation}
which yields
\begin{equation}\label{dSisotropic}
\dd s^2=\frac{\ell^2}{g_+^2}\left\{-g_-^2\dd t^2+\lambda^2\left[\dd \rho^2+\rho^2\left(\frac{4\dd\xi^2}{2-\xi^2}+(1-\xi^2)^2\dd\phi^2\right)\right]\right\}\;,
\end{equation}
where
\begin{equation}
g_\pm=1\pm\frac{\lambda^2\rho^2}{4}\;.
\end{equation}
In these coordinates, the de Sitter horizon has a constant temperature of $T_c=1/(2\pi)$.  $\lambda$ is a gauge parameter that merely scales the radial coordinate $\rho$.  There is an origin at $\rho=0$, the de Sitter horizon is located at $\rho=2/\lambda$, there is an axis of symmetry at $\xi=\pm1$ and a $\mathbb Z_2$ symmetry at $\xi=0$.

This form of de Sitter is suggestively close to the Bach-Weyl (Israel-Khan) solution in polar-Weyl form \eqref{IKpolar}.  Aside from some factors of $f$ and $h$ (which approach unity at large $\rho$), the only differences are that de Sitter in isotropic coordinates has an overall conformal factor of $1/g_+^2$ and a factor of $g_-^2$ in the $\dd t^2$ term whose zero defines the de Sitter horizon.  We will make use of these similarities in our construction.

Now to begin engineering a reference metric, we make some slight modifications to the Bach-Weyl solution \eqref{IKxy} and \eqref{IKpolar}:
\begin{align}\label{refG}
\dd s^2_{\mathrm{ref}}&=\frac{\ell^2}{g_+^2}\left\{-fg_-^2\,F\,\dd t^2+\frac{\lambda^2}{m^2 \Delta_{xy}^2}\left[p^2\left(\frac{4\dd x^2}{(2-x^2)\Delta_x}+\frac{4\dd y^2}{(2-y^2)\Delta_y}\right)+y^2(2-y^2)(1-y^2)^2\,s\,\dd \phi^2\right]\right\}\nonumber\\
&=\frac{\ell^2}{g_+^2}\left\{-fg_-^2\,F\,\dd t^2+\frac{\lambda^2h}{f}\left[\dd \rho^2+\rho^2\left(\frac{4\dd\xi^2}{2-\xi^2}+\frac{(1-\xi^2)^2}{h}\,s\,\dd\phi^2\right)\right]\right\}\;.
\end{align}
Here,
\begin{equation}
s=1-\alpha(1-y^2)^2\;,
\end{equation}
where $\alpha$ is a new parameter, $F$ is a complicated function that we will describe later in \eqref{Fdef}, and equality between the first and second lines of~\eqref{refG} (here and in the remainder of this section) is understood to be through the coordinate transformation \eqref{trans}.  We will use $(x,y)$ coordinates in the region near the black holes and inner segment of the axis, and the $(\rho,\xi)$ coordinates near the cosmological horizon.

We only made three changes to the Bach-Weyl (Israel-Khan) solution to arrive at the reference metric~\eqref{refG}. The first is the inclusion of a conformal factor $1/g_+^2$ to facilitate the matching to de Sitter.  The second is a factor of $s(y)$ in the $\dd\phi^2$ term, which we will use to remove the conical singularity in the inner segment of the axis by adjusting the parameter $\alpha$. Concretely, the conical singularity is removed when $\alpha$ takes the value 
\begin{equation}
\alpha =\frac{(1-k)^2 \left(k^2+6 k+1\right)}{(k+1)^4}\,.
\label{eq:alphaspecial}
\end{equation}
The last change is a factor of $g_-^2\,F$ in the $\dd t^2$ term which introduces a cosmological horizon.

We have freedom to choose the function $F$, but the choice is a delicate matter.  For numerical purposes, we wish for $F$ to be smooth in $(x,y)$ or $(\rho,\xi)$ coordinates, depending on where the coordinates are being used.  The DeTurck method also requires that $F$ be chosen to preserve the regularity of both the cosmological horizon and of the black hole horizons \cite{Headrick:2009pv,Wiseman:2011by,Dias:2015nua}.  That is, $F$ must be positive definite and satisfy
\begin{equation}\label{Fhorconds}
F|_{x=\pm1}=\frac{1}{g_-^2}|_{x=\pm 1}\;,\qquad F|_{\rho=2/\lambda}=\frac{h}{f^2}|_{\rho=2/\lambda}\;,
\end{equation}
where we have chosen equality in the above instead of proportionality in order to preserve the de Sitter and black hole temperatures.

In order to make it easier to find a solution in a Newton-Raphson algorithm (see \emph{e.g.} \cite{Dias:2015nua}), we should also choose $F$ to match physical expectations in certain limits.  Specifically, we expect that when the cosmological horizon (at $\rho=2/\lambda$) is large compared to other length scales (\emph{i.e.} $\lambda\ll1$), the spacetime near the cosmological horizon should approach de Sitter~\eqref{dSisotropic} in isotropic coordinates and the spacetime closer to the origin should be approximately described by the Bach-Weyl solution~\eqref{IKxy}  (when $\alpha=0$).  The cosmological horizon is already accommodated by the fact that $f$ and $h$ approach unity for large $\rho$.  That is, by requiring \eqref{Fhorconds}, we already have $F\approx1$ near the cosmological horizon $\rho\approx2/\lambda$ when $\lambda$ is small.

As for near the origin, we add the requirement that
\begin{equation}
F|_{y=0}=\frac{1}{g_-^2}|_{y=0}\;,
\end{equation}
which is consistent with \eqref{Fhorconds}.  If we set $\alpha=0$, then when $\lambda$ and $\rho$ are small, $g_\pm\approx 1$, and the metric \eqref{refG} approaches that of the Bach-Weyl (Israel-Khan) solution~\eqref{IKxy} as desired.

All of these requirements can be satisfied by choosing $F$ to take the form
\begin{equation}\label{Fdef}
F=\frac{G}{f+g_-^2G-fg_-^2G}\;,\qquad \hbox{with }\quad G=\frac{\tfrac{\hat h}{\hat f}(1-x^2)y^2(2-y^2)+g_-^2}{(1-x^2)y^2(2-y^2)+g_-^4}\;,
\end{equation}
where $\hat f$ and $\hat h$ are any smooth, positive definite functions that agree with $f$ and $h$, respectively at $\rho=2/\lambda$.  To choose $\hat f$ and $\hat h$, we first take the expressions for $f$ and $h$ as written in \eqref{frhoz} and \eqref{hrhoz}, and treat them as functions $f(\rho,z)$ and $h(\rho,z)$. We then set $\hat f(\rho,\xi)=f(2/\lambda,\rho\xi\sqrt{2-\xi^2})$ and similarly for $\hat h$.  Note that we cannot use a choice like $\hat f(\rho,\xi)=f(\tfrac{2}{\lambda},\tfrac{2}{\lambda}\xi\sqrt{2-\xi^2})$ as it is not smooth in the $(x,y)$ coordinates at $x=0,y=1$ \cite{Dias:2015nua}.

Now that we have an appropriate Einstein-DeTurck reference metric \eqref{refG}, we can write our metric ansatz for the static binary in de Sitter:
\begin{align}\label{finalG}
\dd s^2 &=\frac{\ell^2}{g_+^2}\Bigg\{-fg_-^2\,F\,\mathcal{T}\,\dd t^2+\frac{\lambda^2}{m^2 \Delta_{xy}^2}\Bigg[w^2\left(\frac{4 \mathcal{A}\,\dd x^2}{(2-x^2)\Delta_x}+\frac{4 \mathcal{B}}{(2-y^2)\Delta_y} \left(\dd y-x\,(1-x^2)\,y\,(2-y^2)(1-y^2)\mathcal{F}\,\dd x \right)^2\right) \nonumber\\
 & \hspace{5.3cm} +y^2(2-y^2)(1-y^2)^2 \,s\, \mathcal{S} \,\dd\phi^2\Bigg]\Bigg\} \\
&=\frac{\ell^2}{g_+^2}\Bigg\{-fg_-^2\,F\,\widetilde{\mathcal{T}}\,\dd t^2+\frac{\lambda^2h}{f}\Bigg[\widetilde{\mathcal{A}}\,\dd \rho^2+\rho^2\Bigg(\frac{4\widetilde{\mathcal{B}}}{2-\xi^2}\left(\dd\xi -\xi \,(2-\xi^2)(1-\xi^2)\,\rho\, \widetilde{\mathcal{F}}\,\dd\rho \right)^2
+\frac{(1-\xi^2)^2}{h}\,s \, \widetilde{\mathcal{S}}\, \dd\phi ^2\Bigg)\Bigg]\Bigg\}\;, \nonumber
\end{align}
where, w.r.t.~\eqref{refG}, we have introduced the unknown functions $\{\mathcal{Q}_j\}\equiv \{ \mathcal{T}, \mathcal{A}, \mathcal{B}, \mathcal{F}, \mathcal{S}\}$ of $(x,y)$ and the unknown functions $\{\tilde{\mathcal{Q}}_j \}\equiv\{ \widetilde{\mathcal{T}}, \widetilde{\mathcal{A}}, \widetilde{\mathcal{B}}, \widetilde{\mathcal{F}}, \widetilde{\mathcal{S}}\}$ of $(\rho,\xi)$. The two coordinate systems is understood to be equivalent under \eqref{trans}. Note that reference metric \eqref{refG} is recovered when we set $\{ \mathcal{T}=1, \mathcal{A}=1, \mathcal{B}=1, \mathcal{F}=0, \mathcal{S}=1\}$  (and similarly for the associated set of tilde functions $\{\tilde{\mathcal{Q}_j}\}$).  

The ansatz \eqref{finalG} is the most general such metric compatible with our symmetries (static and axisymmetric).  The remaining gauge freedom is fixed by solving the Einstein-DeTurck equations, and then afterwards verifying that the solution is not a Ricci soliton.  

Our task is now to find the unknown functions $\{\mathcal{Q}_j\}\equiv \{ \mathcal{T}, \mathcal{A}, \mathcal{B}, \mathcal{F}, \mathcal{S}\}$ and $\{\tilde{\mathcal{Q}}_j \}\equiv\{ \widetilde{\mathcal{T}}, \widetilde{\mathcal{A}}, \widetilde{\mathcal{B}}, \widetilde{\mathcal{S}}, \widetilde{\mathcal{F}}\}$ in \eqref{finalG} by solving the Einstein-De Turck equations subject to the appropriate physical boundary conditions. All boundary conditions are determined by symmetry or regularity.  A detailed discussion of regularity boundary conditions can be found in in Section V of the review \cite{Dias:2015nua}.  Here, we just mention that we have chosen several factors of $x$, $1-x$, $y$ and $1-y$ (and similarly for $\xi,\rho$) in \eqref{refG} and \eqref{finalG} so that regularity is automatically enforced if the functions are finite.  More specific conditions on the functions can then be derived from a series expansion of the equations of motion.  

It turns out that on all boundaries (except for one), we impose homogeneous Neumann conditions for all functions in their respective coordinate patches.  So, for instance at the $\mathbb{Z}_2$ symmetry about the axis $x=0$, we impose $\partial_x \mathcal{Q}_j|_{x=0}=0$.  At the black hole horizon at $x=1$, we similarly impose $\partial_x \mathcal{Q}_j|_{x=1}=0$, and so on for all the other boundaries, with one exception.  At $\rho=2/\lambda$ we have the cosmological horizon with temperature $T_c=1/(2\pi)$. Regularity at this boundary requires that we set $ \widetilde{\mathcal{T}}|_{\rho=2/\lambda}=\widetilde{\mathcal{A}}|_{\rho=2/\lambda}$,  $ \widetilde{\mathcal{F}}|_{\rho=2/\lambda}=0$ and Robin (mixed) boundary conditions for the other functions (we have an arbitrary choice of whether to impose a mixed Robin condition on $\widetilde{\mathcal{T}}$ or on $\widetilde{\mathcal{A}}$).  These conditions simply follow from the equations of motion, and that it is not too enlightening to explicitly display them here \cite{Dias:2015nua}.

Because we work with a metric in two separate coordinate systems, we will need to add some means of consistently combining the coordinate patches together.  We handle this using a set of ``patching conditions".  The details of this process are given later in section \ref{sec:patching}.

The solutions we seek are parametrized by $\alpha$, $\lambda$ and $k$.  On the desired solution that is free of conical singularities, $\alpha$ is fixed according to \eqref{eq:alphaspecial}. The coordinate location of the cosmological horizon, $\rho=2/\lambda$, is fixed by a choice of $\lambda$, but the horizon has a constant temperature of $T_c=1/(2\pi)$. On physical grounds, the static de Sitter binary is a 1-parameter family of solutions parametrized by $T_+/T_c$.  Therefore, any combination of $k$ and $\lambda$ that give the same black hole temperature $T_+$ are physically equivalent.  To collect our numerical data, we typically have fixed $\lambda=1/10$ and used $k$ to parametrize our solutions.  We have tried different values of $\lambda$, but after trial and error, this value typically generated the best numerical results.

\section{Patching and Numerical Methods}\label{sec:patching}
In this section we explain how we partition the domain of integration using patching techniques (see \emph{e.g.} \cite{Dias:2015nua}). As mentioned in the main text, the solution we seek contains five boundaries: the inner segment of the axis ($\partial_{\phi}^{\rm in}$), the black hole horizon ($\mathcal{H}^+$), the outer segment of the axis ($\partial_{\phi}^{\rm out}$), the cosmological horizon ($\mathcal{H}^c$) and the plane of $\mathbb{Z}_2$ symmetry.

\begin{figure}
    \centering
    \includegraphics[width=0.98\textwidth]{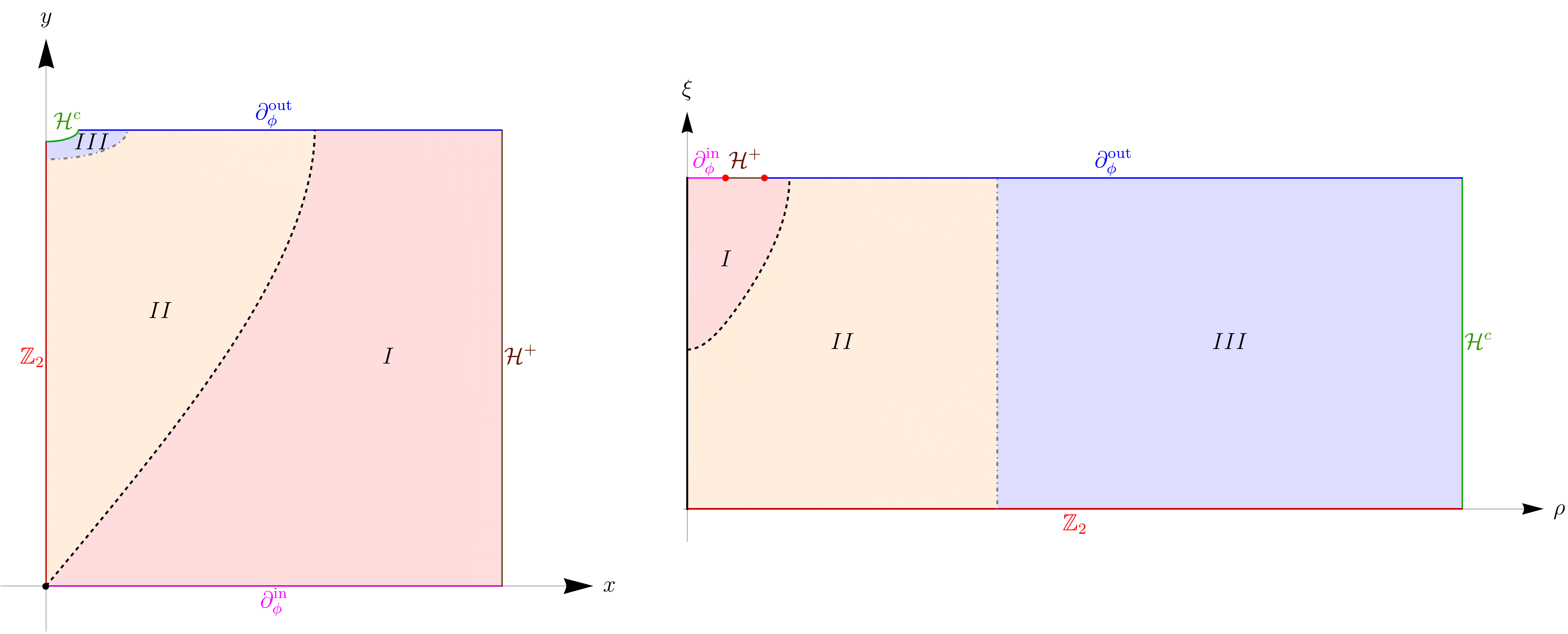}
    \caption{Illustration of the patches used in our numerical construction.}
    \label{fig:map}
\end{figure}

We numerically solve the Einstein-DeTurck equations with the boundary conditions of regularity at each of our five boundaries.  To do so, we use both the coordinates $(\rho,\xi)$ and $(x,y)$ of the preceding section. Near the black hole event horizon we use $(x,y)$ coordinates, while near the cosmological horizon we use $(\rho,\xi)$ coordinates. We used a total of three patches $-$ $I$, $II$ and $III$ $-$ with each of these having four boundaries: see FIG.~\ref{fig:map}. Patches $I$ and $II$ are defined in $(x,y)$ coordinates, and patch $III$ in $(\rho,\xi)$ coordinates. The patching boundary between the patch $I$ and $II$ (dashed line in FIG.~\ref{fig:map}) is given by $x=x_0 y\sqrt{2-y^2}$, with $y\in(0,1)$. The patching boundary between patch $II$ and patch $III$ is simply given by $\rho=\rho_0$.

Furthermore, we fix $\rho_0$ and $x_0$ through
\begin{equation}
\rho_0=\frac{2}{3} \left(\frac{2}{\lambda }-\frac{1}{k}\right)+\frac{1}{k}\quad\text{and}\quad x_0=\frac{1}{2} \left(1-\sqrt{1-\sqrt{1-\frac{1}{k^2 \rho _0^2}}}\right)+\sqrt{1-\sqrt{1-\frac{1}{k^2 \rho _0^2}}}\,,
\end{equation}
so that the only free parameters are $\alpha$, $\lambda$, and $k$. Note that for non-singular solutions we require $\alpha$ to be given by \eqref{eq:alphaspecial} and, as described in the end of the previous section, we fix $\lambda=1/10$ for numerical convenience.  We now apply the numerical methods detailed in \cite{Dias:2015nua}, and discretize each of our patches on a $N\times N$ Chebyshev-Gauss-Lobatto grid using transfinite interpolation and pseudospectral collocation, for a total grid size of $(N+N+N)\times N$.  The discretized Newton-Raphson equation reduces to an iteration of linear matrix problems, which we solve by LU decomposition. To find the first solution (which is always the trickiest step of the Newton-Raphson method), we have set $\lambda = 1/10$, $k=1/2$ and used a Newton-Raphson algorithm with a judiciously chosen damping factor.  Additionally, we used what we call the ``$\delta$-trick" as explained in section VII.A of \cite{Dias:2015nua}.

\begin{figure}[ht]
\centering
\includegraphics[width=0.45\textwidth]{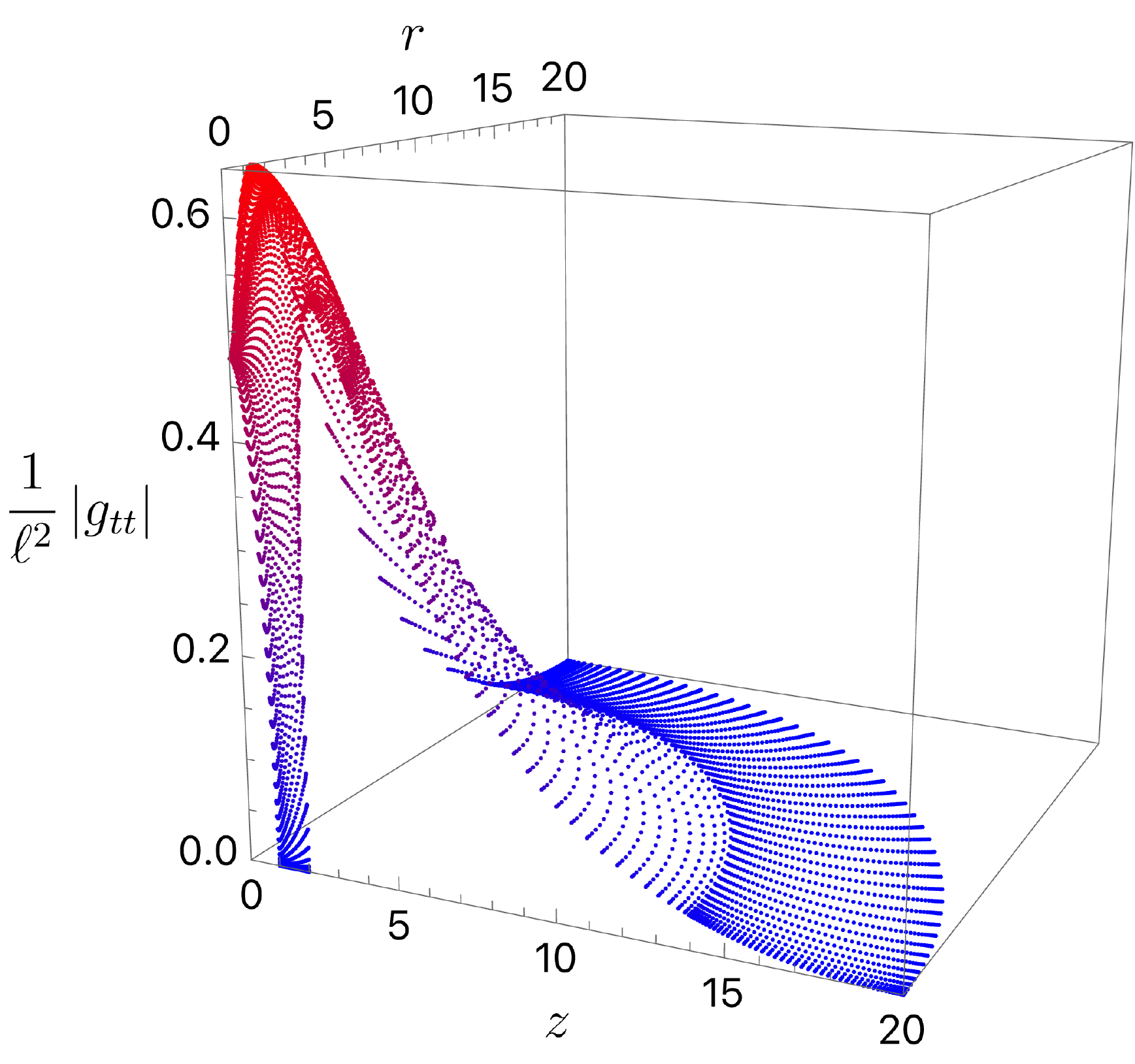}
\includegraphics[width=0.45\textwidth]{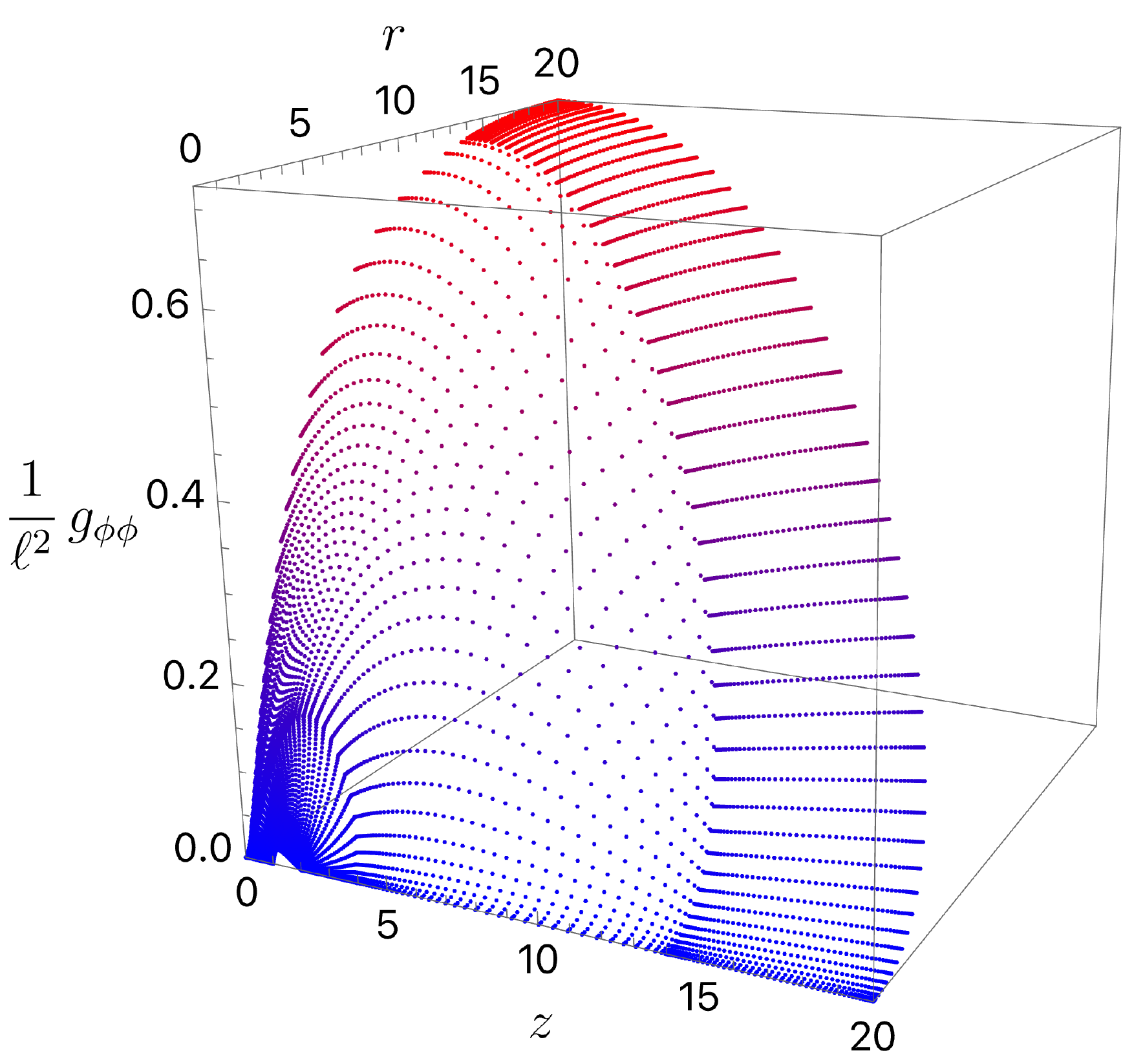}
\caption{The gauge invariant metric functions $-g_{tt}$ and $g_{\phi\phi}$ for $T_+/T_c=3.75$ ($\lambda=0.1, k=0.5$).}\label{fig:functionsk05}
\end{figure}

\begin{figure}[ht]
\centering
\includegraphics[width=0.45\textwidth]{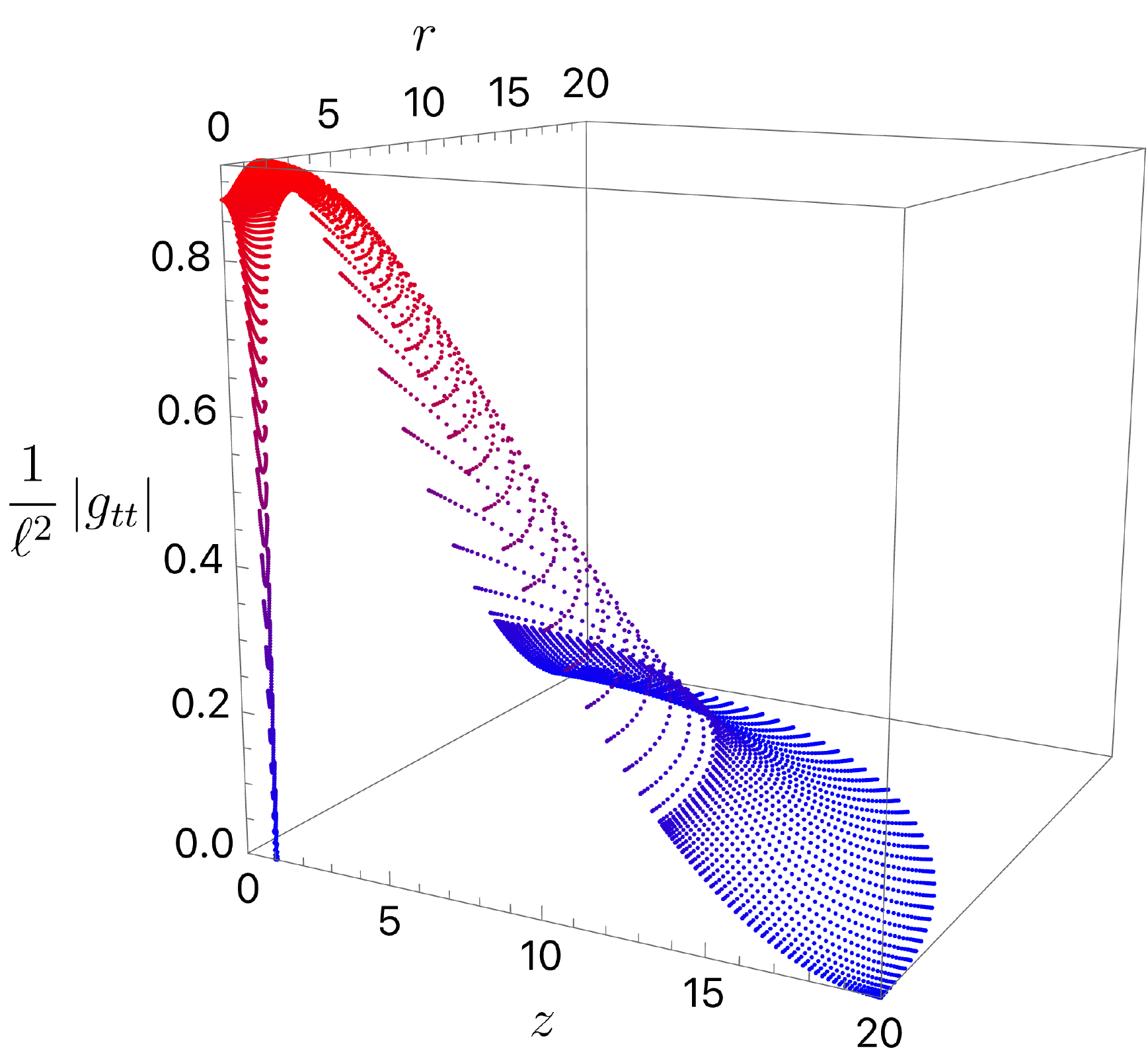}
\includegraphics[width=0.45\textwidth]{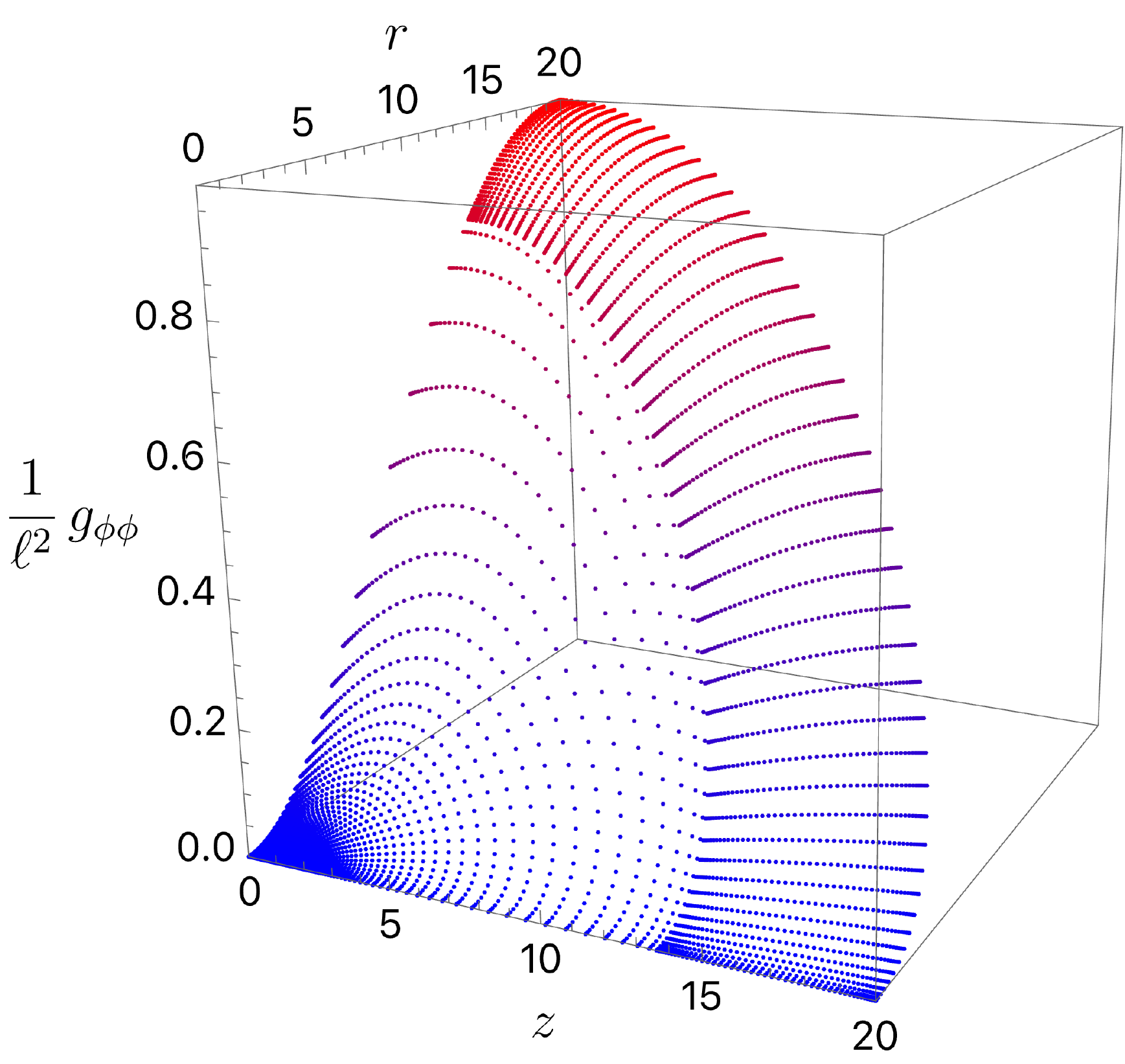}
\caption{The gauge invariant metric functions $-g_{tt}$ and $g_{\phi\phi}$ for $T_+/T_c=63.60$ ($\lambda=0.1, k=0.9295$).}\label{fig:functionsk09295}
\end{figure}

To illustrate how our solutions look like, (besides FIG.~3 in the main text) in FIGs.~\ref{fig:functionsk05}$-$\ref{fig:functionsk09295} we display the two metric functions $|g_{tt}|$ and $g_{\phi\phi}$ $-$ that are gauge invariant (since $\partial_t$ and $\partial_\phi$ are Killing vector fields) $-$  as a function of the original cylindrical Weyl coordinates $(r,z)$ for two representative solutions with black hole temperature $T_+/T_c=3.75$ (FIG.~\ref{fig:functionsk05}) and $T_+/T_c=63.60$ (FIG.~\ref{fig:functionsk09295}). Recall that in this coordinate chart we have a ``rod structure" where the rotation axis and the black hole horizons are all located at $r=0$. More concretely, at $r=0$, the black horizons lie in the regions $z\in (1,1/k)$ and $z\in (-1/k,-1)$ and FIGs.~\ref{fig:functionsk05}$-$\ref{fig:functionsk09295} indeed show that $g_{tt}$ vanishes in these segments. Moreover, still at $r=0$, the inner segment of the axis between the black holes is in the region $z\in(-1,1)$ and the outer segments of the axis are in $z\in(1/k,\infty)$ and $z\in(-\infty,-1/k)$: the right panels of FIGs.~\ref{fig:functionsk05}$-$\ref{fig:functionsk09295} indeed show that $g_{\phi\phi}$ vanishes in these segments (in  FIG.~\ref{fig:functionsk09295} $g_{\phi\phi}$ does not vanish in the segment $1<z<1/k$, but this is difficult to see in the image). The solution has a cosmological horizon at $\sqrt{r^2+z^2}=2/\lambda$ (with temperature $T_c=1/(2\pi)$) where $g_{tt}$ also vanishes as clearly identified in the left panels of FIGs.~\ref{fig:functionsk05}$-$\ref{fig:functionsk09295}. The system is $\mathbb{Z}_2$ symmetric about $z=0$, i.e. $g_{ab}(r,-z)=g_{ab}(r,z)$ and thus we just display the solution for $z\geq 0$. The boundary $\sqrt{r^2+z^2}=\rho_0$ that marks the smooth transition between patches II and III is very clearly identified in FIGs.~\ref{fig:functionsk05}$-$\ref{fig:functionsk09295} (of course, the patching between patches I and II is also smooth; it is harder to identify it by just looking at the plots).

\section{Convergence Tests}
In this section, we show that the norm $\chi\equiv \xi^a \xi_a$ of the DeTurck vector vanishes in the continuum limit, as expected for a solution of the Einstein-DeTurck equation that is not a Ricci soliton (\emph{i.e.} that is instead a true solution to the Einstein equation). Additionally, we find  exponential convergence, which is consistent with the use of pseudospectral collocation methods.

\begin{figure}
    \centering
    \includegraphics[width=0.49\textwidth]{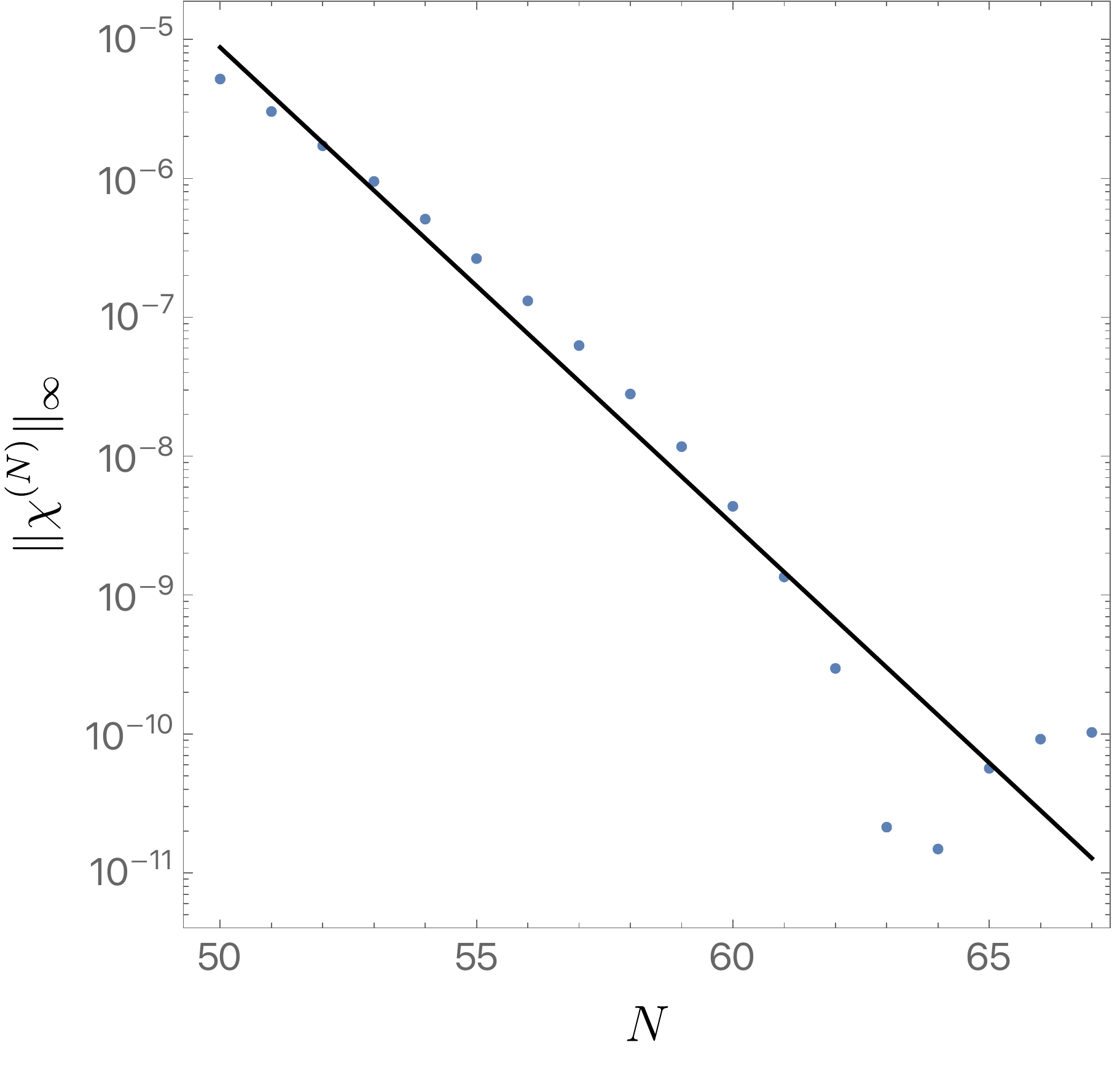}
    \caption{Convergence test showing both the exponential accuracy of our numerical method and the fact that we are not converging to a Ricci soliton.}
    \label{fig:convergence}
\end{figure}

Let $\chi^{(N)}$ be $\chi$ computed on a (3-patched) grid with $(N+N+N)\times N$ spectral collocation points. For concreteness we take $k=1/2$, $\lambda=1/10$ and $\alpha$ as given in~\eqref{eq:alphaspecial}. In FIG.~\ref{fig:convergence} we show $\lVert \chi^{(N)}\rVert_{\infty}$ as a function of $N$ in a $\log$-plot. The solid black line shows the best $\chi^2$-fit to a straight line in the $\log$-plot, and yields
\begin{equation}
f(N)=27.89409 - 0.79076\,N\,.
\end{equation}
The exponential trend is clear and confirms that the Einstein-DeTurck solution is converging to a true solution of the Einstein equation (and not to a Ricci soliton).

\bibliographystyle{utphys-modified}
\bibliography{papers}
\end{document}